\documentclass[preprint,12pt]{elsarticle}

\usepackage{graphicx}

\usepackage{amssymb}

\journal{Physics Letters A}

\begin{document}

\begin{frontmatter}



\title{Analytical solutions for the radial Scarf II potential}


\author[atomki]{G. L\'evai\corref{cor1}\fnref{telfax}}
\ead{levai@atomki.mta.hu}

\address[atomki]{
Institute for Nuclear Research, Hungarian Academy of Sciences 
(Atomki), Debrecen, Pf.~51, Hungary 4001}

\author[ud]{\'A. Baran}

\address[ud]{Faculty of Informatics, University of Debrecen, Debrecen, 
         Pf.~400, Hungary 4002}
         
\author[atomki]{P. Salamon} 

\author[atomki,ud]{T. Vertse}

\cortext[cor1]{Corresponding author}
\fntext[telfax]{Tel: +36-52-509238, Fax: +36-52-416181}

\begin{abstract}
The real Scarf II potential is discussed as a radial problem. This 
potential has been studied extensively as a one-dimensional problem, 
and now these results are used to construct its bound and resonance 
solutions for $l=0$ by setting the origin at some arbitrary value 
of the coordinate. The solutions with appropriate boundary conditions 
are composed as the linear combination of the two independent 
solutions of the Schr\"odinger equation. The asymptotic expression 
of these solutions is used to construct the $S_0(k)$ $s$-wave 
$S$-matrix, the poles of which supply the $k$ values corresponding 
to the bound, resonance and anti-bound solutions. The location of 
the discrete energy eigenvalues is analyzed, and the relation of 
the solutions of the radial and one-dimensional Scarf II potentials 
is discussed. It is shown that the generalized Woods--Saxon potential 
can be generated from the Rosen--Morse II potential in the same way 
as the radial Scarf II potential is obtained from its one-dimensional  
correspondent. Based on this analogy, possible applications are also 
pointed out. 
\end{abstract}


\end{frontmatter}



\section{Introduction}

Exactly solvable quantum mechanical potentials proved to be invaluable tools 
in the understanding of many fundamental quantum mechanical concepts. In 
particular, they give insight into complex phenomena, like the symmetries 
of quantum mechanical systems, and they allow the investigation of 
transitions through critical parameter domains. Besides this, analytical 
solutions serve as a firm basis for the development of numerical techniques. 

The one-dimensional Schr\"odinger equation 
\begin{equation}
-\psi''(x)+V(x)\psi(x)=E\psi(x) 
\label{scheq}
\end{equation}
occurs in many applications. Here the potential function and the energy 
eigenvalue are defined such that they contain reduced mass $m$ and $\hbar$ as 
$V(x)=2mv(x)/\hbar^2$ and $E=2m \epsilon/\hbar^2$, so their physical dimension 
is distance$^{-2}$. In the simplest case (\ref{scheq}) is defined on the 
full $x$ axis, i.e. $x\in(-\infty,\infty)$, while for spherical 
potentials defined in higher, typically three dimension, Eq.  
(\ref{scheq}) can be obtained after the separation of the angular 
variables, if only the $s$-wave ($l=0$) solutions are considered. In 
this case the problem is defined on the positive half axis, $r\in[0,\infty)$, 
and the $x$ variable is denoted by $r$. Besides these options, (\ref{scheq}) 
can also be defined on finite sections of the real $x$ axis, or even on 
more complicated trajectories of the complex $x$ plane, but we shall not 
consider these in the present work. 

Being a second-order ordinary differential equation, (\ref{scheq}) has 
two independent solutions, and the physical solutions can be obtained 
as linear combination of these, satisfying the appropriate boundary 
conditions. Due to normalizability, bound states have to vanish at the 
boundaries (i.e. $x=\pm\infty$ in one dimension, and $r=0$ and $r=\infty$ 
in the radial case). Unbound solutions, e.g. scattering and resonance 
solutions also have to satisfy asymptotic boundary conditions, depending 
on the nature of the potential. If $V(x)$ vanishes exactly or exponentially 
for $x\rightarrow\pm\infty$, then 
these solutions of the one-dimensional problem have exponential 
asymptotic components $\exp(\pm {\rm i}kx)$, where $E=k^2$. In the 
radial case the same asymptotics are valid for $r\rightarrow\infty$, 
while for $r=0$ these solutions have to vanish. 

There are some potentials that are defined both as one-dimensional 
and as radial problems, e.g. the harmonic oscillator. The bound-state 
solutions of these two problems are related to each other in a special 
way: the odd wave functions of the one-dimensional potential, which vanish 
at $x=0$, are identical for $x\ge 0$ to the $s$-wave ($l=0$) radial 
wave functions, and the energy eigenvalues are also identical. 

A rather effective way for the unified discussion of bound, scattering 
and resonance solutions 
in asymptotically vanishing potentials 
is the application of the transmission coefficient 
$T(k)$ (in one dimension) and the $s$-wave $S$ matrix $S_0(k)$ (in the 
radial case). These quantities can be constructed from the asymptotic 
solutions, and their poles correspond to the bound, anti-bound and 
resonance states. From the exact solutions of these problems $T(k)$ 
and $S_0(k)$ can also be expressed in closed analytic form. 

Here we dicuss the Scarf II potential as a radial problem. This 
potential has two independent terms and 
belongs to the shape-invariant \cite{gendenshtein} subclass 
of the Natanzon potential class \cite{natanzon}, which contains 
problems with bound-state solutions written in terms of a single 
hypergeometric function. The first reference to potential 
(\ref{sciipot}) in the English literature occurred in 1983 in  
Ref. \cite{gendenshtein}, so it is sometimes referred to as the 
Gendenshtein potential. However, it was already mentioned a year 
before in a Russian monograph \cite{natpriv}. Its detailed 
description was presented later, e.g.  
the normalization coefficients of its bound-state solutions 
have been calculated only recently \cite{pla02}. The transmission 
and reflection coefficients have been given in Ref. \cite{ks88}, with 
corrections added in Ref. \cite{jpa01}. It has been a favourite toy 
model in ${\cal PT}$-symmetric quantum mechanics, where it was used 
to demonstrate the breakdown of ${\cal PT}$ symmetry 
\cite{Ahm01a,mpla01a}. Further studies concerned its algebraic 
\cite{BQ00,jpa01} and scattering aspects \cite{jpa01}, the combined 
effects of SUSYQM and ${\cal PT}$ symmetry \cite{jpa02b}, the 
pseudo-norm of its bound states \cite{pla02}, the handedness (chirality) 
effects in scattering \cite{Ahm04}, spectral singularities \cite{Ahm09}, 
unidirectional invisibility \cite{Ahm13} and the accidental crossing 
of its energy levels \cite{Ahm15}. 

Despite its prominent status as a one-dimensional quantum system, the 
Scarf II potential has not been considered yet as a radial problem. Here we 
fill this gap by introducing a lower cut at a certain $x=r_0$ value 
and prescribing the appropriate boundary conditions. We construct 
the $S$-matrix for the $s$-wave solutions, $S_0(k)$, and determine 
its poles on the complex $k$ plane to identify its bound, anti-bound 
and resonance solutions. This will be done in Sec. \ref{radscii}, 
following the discussion of the one-dimensional problem for reference 
in Sec. \ref{1dscii}. In Sec. \ref{gwsanalogy} the analogy with the 
case of the generalized Woods--Saxon and the Rosen--Morse II 
potentials will be outlined, and possible applications are pointed out. 
Finally the results are summarized in Sec. \ref{summary}.

\section{The Scarf II potential in one dimension}
\label{1dscii}

A possible parametrization of this potential is \cite{jpa02b} 
\begin{equation}
V(x)=-\frac{V_1}{\cosh^2(cx)}
+\frac{V_2\sinh(cx)}{\cosh^2(cx)}\ ,
\label{sciipot}
\end{equation}
where
\begin{equation}
V_1=c^2\left(\frac{\alpha^2+\beta^2}{2}-\frac{1}{4}\right)\hspace{2cm} 
V_2={\rm i}c^2\frac{\beta^2-\alpha^2}{2}\ , 
\label{v1v2}
\end{equation}
and $c>0$ is a scaling factor of the coordinate.  
This potential is real if $\alpha^*=\beta$ holds, while it is 
${\cal PT}$-symmetric if $\alpha$ and $\beta$ are real or 
imaginary. In what follows we consider the real version only. 
Potential (\ref{sciipot}) is depicted in Fig. \ref{fig1} for 
some values of the parameters. It has a minimum $x_-$ and a 
maximum $x_+$ at 
\begin{equation}
x_{\pm}=c^{-1}\sinh^{-1}\left[\frac{V_1}{V_2}
 \pm \left[\left(\frac{V_1}{V_2}\right)^2+1\right]^{1/2}\right]\ .
\label{xpxm}
\end{equation}
The potential reflected by $x=0$ can be constructed easily by 
considering $V_2\rightarrow -V_2$, i.e. 
$\alpha\leftrightarrow\beta$. 

The bound-state wave functions are 
\begin{equation}
\psi_n(x) =C_n
(1-{\rm i}\sinh(cx))^{
 \frac{\alpha}{2}+\frac{1}{4}}
 (1+{\rm i}\sinh(cx))^{\frac{\beta}{2}+\frac{1}{4}}
P_n^{(\alpha,\beta)}({\rm i}\sinh(cx))\ ,
\label{sciiwf}
\end{equation}
while the corresponding energy eigenvalues are written as 
\begin{equation}
E_n=
 -c^2\left(n+\frac{\alpha+\beta+1}{2}\right)^2\ . 
\label{sciien}
\end{equation}
Normalizability of (\ref{sciiwf}) requires 
\begin{equation}
n<-\frac{1}{2}[{\rm Re}(\alpha+\beta)+1]\ .
\label{sciincond}
\end{equation}
$C_n$ in (\ref{sciiwf}) was calculated for the real and the 
${\cal PT}$-symmetric 
version of the Scarf II potential in Ref. \cite{pla02}. In the former 
case it can be written as 
\begin{equation}
C_n=2^{-\frac{\alpha+\beta}{2}-1}
\left[c\frac{\Gamma(-\alpha-n)\Gamma(-\beta-n)(-\alpha-\beta-2n-1)n!
}{
\Gamma(-\alpha-\beta-n)\pi
}
\right]^{1/2}\ . 
\label{Cn}
\end{equation}
Note that although $\alpha$ and $\beta$ are complex, $C_n$ is real due to 
$\alpha=\beta^*$ and Eq. (\ref{sciincond}). It can also be proven that 
the bound-state wave functions (\ref{sciiwf}) are real for even $n$, and 
imaginary for odd $n$: this can be demonstrated by expressing the 
complex conjugate of (\ref{sciiwf}), which turns out to be 
$[\psi_n(x)]^*=(-1)^n \psi_n(x)$, due to Eq. 22.4.1 of 
Ref. \cite{AS70}. 

With a reparametrization, the notation of Ref. \cite{jpa01} can 
be obtained, in which the scattering aspects of the one-dimensional 
Scarf II potential have been discussed. Taking 
\begin{equation}
\alpha=-s-\frac{1}{2}-{\rm i}\lambda\ , \hspace{1cm} 
 \beta=-s-\frac{1}{2}+{\rm i}\lambda
\label{ab}
\end{equation}
one obtains 
\begin{eqnarray}
V_1=c^2[s(s+1)-\lambda^2] \hspace{1cm} V_2=c^2(2s+1)\lambda
\label{v1v2sl}
\end{eqnarray}
in Eq. (\ref{sciipot}). 
According to (\ref{ab}), the Scarf II potential will be real for real 
values of $s$ and $\lambda$. 
Note that this potential remains invariant if the signs of $s+1/2$ and 
$\lambda$ are reversed simutaneously. This means that without the loss 
of generality one can require $s>-1/2$. Condition (\ref{sciincond}) is now 
$n<{\rm Re}(s)=s$, so in order to obtain normalizable states one needs 
$s>0$. It is notable that $E_n$ in (\ref{sciien}) depends only on 
$s=-(\alpha+\beta+1)/2$, and is independent of 
$\lambda={\rm i}(\alpha-\beta)/2$. 

In Ref. \cite{jpa01} the general solutions of the Schr\"odinger equation with 
the potential (\ref{sciipot}) and (\ref{v1v2sl}) are expressed in terms of 
hypergeometric functions as
\begin{eqnarray}
F_1(x) &&= 
(1-{\rm i}\sinh(cx))^{-\frac{s+{\rm i}\lambda}{2}}
 (1+{\rm i}\sinh(cx))^{-\frac{s-{\rm i}\lambda}{2}}
\nonumber\\
 &&\times 
 _2F_1(-s-{\rm i}k, -s+{\rm i}k; {\rm i}\lambda-s+1/2; (1+{\rm i}\sinh(cx))/2) 
\label{sol1}
\end{eqnarray}
and 
\begin{eqnarray}
F_2(x) &&= 
A(1-{\rm i}\sinh(cx))^{-\frac{s+{\rm i}\lambda}{2}}
 (1+{\rm i}\sinh(cx))^{\frac{s+1-{\rm i}\lambda}{2}}
\nonumber\\
 &&\times _2F_1(1/2-{\rm i}\lambda-{\rm i}k, 1/2-{\rm i}\lambda+{\rm i}k; 
  s+3/2-{\rm i}\lambda; (1+{\rm i}\sinh(cx))/2)\ .
\nonumber\\
\label{sol2}
\end{eqnarray}
Note that (\ref{sol2}) is obtained 
from (\ref{sol1}) by Eq. 15.5.4 of Ref. \cite{AS70}, 
where $A=2^{{\rm i}\lambda -s -1/2}$. Note also that the two solutions 
are connected by the $s+1/2\leftrightarrow {\rm i}\lambda$ transformation, 
which leaves (\ref{v1v2sl}), and thus (\ref{sciipot}) invariant. 

The asymptotic behavior of (\ref{sol1}) and (\ref{sol2}) can be obtained by 
applying 15.3.4, 15.3.5 and 15.3.6 of Ref. \cite{AS70}, and the results are 
\begin{eqnarray}
\lim_{x\rightarrow\infty} F_1(x) &=&  
a_{1+}\exp({\rm i}kx)+b_{1+}\exp(-{\rm i}kx) 
\label{ab1+}\\ 
\lim_{x\rightarrow -\infty} F_1(x) &=& 
a_{1-}\exp({\rm i}kx)+b_{1-}\exp(-{\rm i}kx) 
\label{ab1-}\\
\lim_{x\rightarrow\infty} F_2(x) &=& 
a_{2+}\exp({\rm i}kx)+b_{2+}\exp(-{\rm i}kx) 
\label{ab2+}\\
\lim_{x\rightarrow -\infty} F_2(x) &=& 
a_{2-}\exp({\rm i}kx)+b_{2-}\exp(-{\rm i}kx)\ ,
\label{ab2-}
\end{eqnarray}
where 

\begin{eqnarray} 
a_{1+}=D_1 2^{-s-2{\rm i}k/c}{\rm e}^{\pi(k/c-\lambda-{\rm i}s)/2} && 
b_{1+}=C_1 2^{-s+2{\rm i}k/c}{\rm e}^{\pi(-k/c-\lambda-{\rm i}s)/2} 
\label{ab1p}\\
a_{1-}=b_{1+}{\rm e}^{\pi(k/c+\lambda+{\rm i}s)} && 
b_{1-}=a_{1+}{\rm e}^{\pi(-k/c+\lambda+{\rm i}s)} 
\label{ab1m}\\
a_{2+}=D_2 2^{-s-2{\rm i}k/c}{\rm e}^{\pi(k/c+\lambda+{\rm i}(s+1))/2} 
\hspace{.4cm} && 
b_{2+}=C_2 2^{-s+2{\rm i}k/c}{\rm e}^{\pi(-k/c+\lambda+{\rm i}(s+1))/2} 
\nonumber\\
\label{ab2p}\\
a_{2-}=b_{2+}{\rm e}^{\pi(k/c-\lambda-{\rm i}(s+1))} && 
b_{2-}=a_{2+}{\rm e}^{\pi(-k/c-\lambda-{\rm i}(s+1))}\ , 
\label{ab2m}
\end{eqnarray}
and 

\begin{eqnarray}
C_1=\frac{\Gamma({\rm i}\lambda+1/2-s)\Gamma(-2{\rm i}k/c)}{
 \Gamma({\rm i}\lambda+1/2-{\rm i}k/c)\Gamma(-s-{\rm i}k/c)}
 &&
D_1=\frac{\Gamma({\rm i}\lambda+1/2-s)\Gamma(2{\rm i}k/c)}{
 \Gamma({\rm i}\lambda+1/2+{\rm i}k/c)\Gamma(-s+{\rm i}k/c)}
 \nonumber\\
\label{cd1}\\
C_2=\frac{\Gamma(-{\rm i}\lambda+3/2+s)\Gamma(-2{\rm i}k/c)}{
 \Gamma(s+1-{\rm i}k/c)\Gamma(-{\rm i}\lambda+1/2-{\rm i}k/c)}
 \hspace{.4cm}
 &&
D_2=\frac{\Gamma(-{\rm i}\lambda+3/2+s)\Gamma(2{\rm i}k/c)}{
 \Gamma(s+1+{\rm i}k/c)\Gamma(-{\rm i}\lambda+1/2+{\rm i}k/c)}
 \nonumber\\
\label{cd2}
\end{eqnarray}

For a wave traveling to the right the transmission and reflection 
coefficients are expressed as \cite{ks88,jpa01}

\begin{eqnarray}
T(k)&=&\frac{a_{1+}b_{2+}-b_{1+}a_{2+}}{a_{1-}b_{2+}-b_{1+}a_{2-}} 
\nonumber\\
&=&
\frac{\Gamma(-s-{\rm i}k/c)\Gamma(s+1-{\rm i}k/c)
      \Gamma({\rm i}\lambda+1/2-{\rm i}k/c)\Gamma(-{\rm i}\lambda+1/2-{\rm i}k/c)
}{\Gamma(-{\rm i}k/c)\Gamma(1-{\rm i}k/c)\Gamma^2(1/2-{\rm i}k/c)}
\nonumber\\
\label{tk}
\\
R(k)&=&\frac{b_{1-}b_{2+}-b_{1+}b_{2-}}{a_{1-}b_{2+}-b_{1+}a_{2-}} 
\nonumber\\
&=&
T(k)\left(\frac{\cos(\pi s)\sinh(\pi\lambda)}{\cosh(\pi k/c)}
   +{\rm i}\frac{\sin(\pi s)\cosh(\pi\lambda)}{\sinh(\pi k/c)}\right)\ .
\nonumber\\
\label{rk}   
\end{eqnarray}

The poles of $T(k)$ are located at $-n=-s-{\rm i}k/c$, $-n=s+1-{\rm i}k/c$, 
$-n=-{\rm i}\lambda+1/2-{\rm i}k/c$ and $-n={\rm i}\lambda+1/2-{\rm i}k/c$. 
The first choice corresponds to the energy eigenvalues 
\begin{equation}
E_n=-c^2(s-n)^2
\label{sciiensl}
\end{equation}
in accordance with (\ref{sciincond}), and converts $F_1(x)$ in (\ref{sol1}) 
into (\ref{sciiwf}) (up to the constant factor 
$(-1)^n n! \Gamma(\beta+1)[\Gamma(\beta+n+1)C_n]^{-1}$) after applying 
Eqs. 15.3.6 and 22.5.42 of Ref. \cite{AS70}. 
The second one stands for anti-bound or virtual states with $k$ 
located on the negative imaginary axis, while the last two poles 
correspond to non-normalizable 
complex-energy states, i.e. resonances with 
$E_n=k^2=-c^2(n-1/2\pm {\rm i}\lambda)^2$.

\section{The Scarf II potential as a radial problem}
\label{radscii}

\begin{figure}[h]
\begin{center}
\resizebox*{10cm}{!}{{\includegraphics{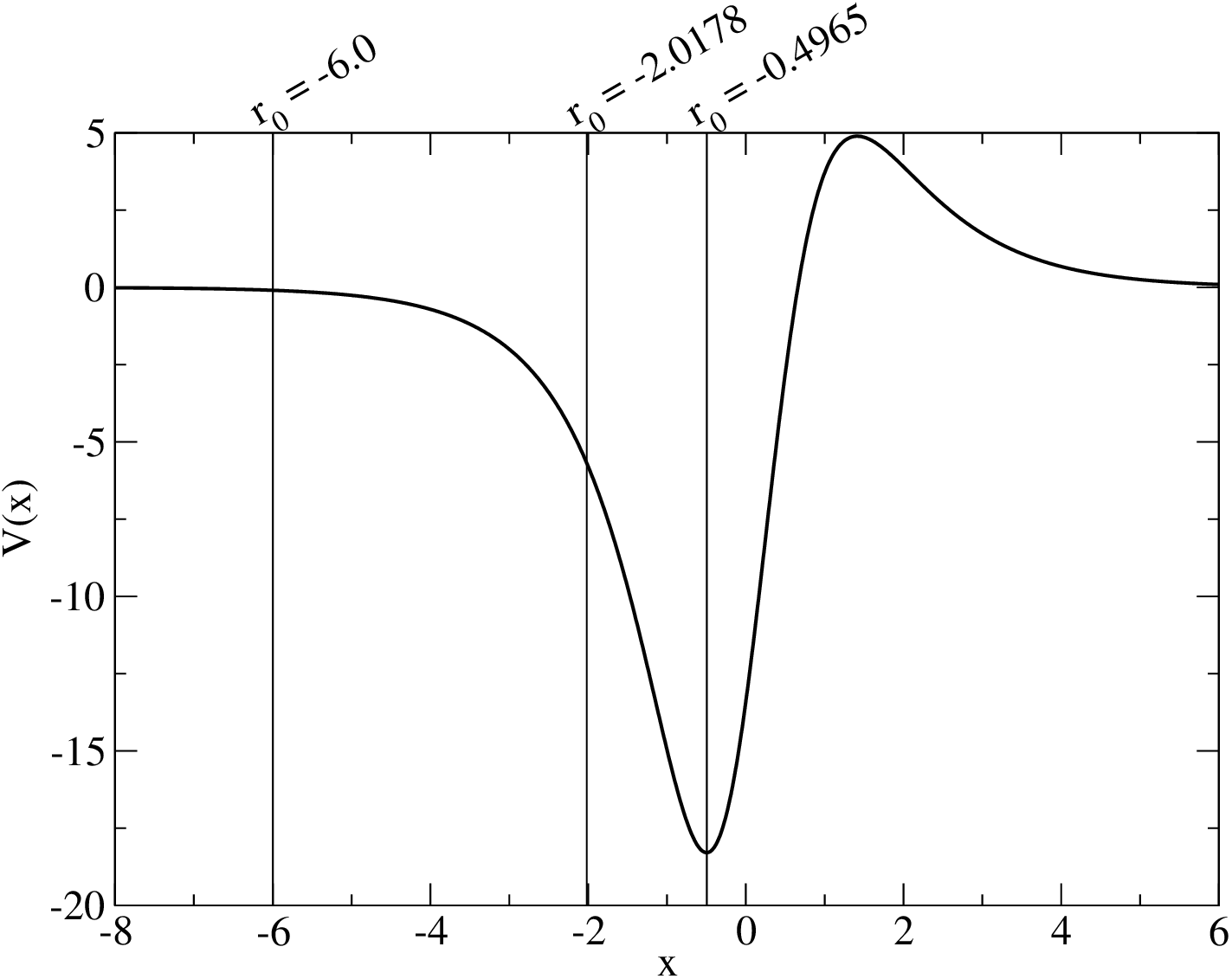}}} 
\end{center}
\caption{The structure of the one-dimensional Scarf II 
potential for $V_1=13.4$ and $V_2=18.92$ corresponding to 
$c=1$, 
$\alpha=-4.3-2.2{\rm i}=\beta^*$, $s=3.8$ and $\lambda=2.2$. 
This potential has four bound states 
at $E_0=-14.44$, $E_1=-7.84$, $E_2=-3.24$ and $E_3=-0.64$. 
The vertical lines at $x=r_0=-0.4965$, $-2.017825$ (abbreviated in the 
plot) and $-6.0$   
define three radial potentials (see Sec. \ref{radscii}) with different  
location of the origin. In the first case $r_0=x_-$, while the 
second $r_0$ corresponds to the first node of $\psi_2(x)$ in 
(\ref{sciiwf}). 
}
\label{fig1}
\end{figure}

In this case the general wave function is constructed from the linear 
combination of the two independent solutions (\ref{sol1}) and (\ref{sol2}) 
with boundary condition that it should vanish at the origin. The position 
of the origin need not be chosen at $x=0$, rather one can cut the 
one-dimensional potential (\ref{sciipot}) at an arbitrary finite 
value. Let us thus define $x=r+r_0$, where $r\in[0,\infty)$, i.e. 
$x\in [r_0,\infty)$. Figure \ref{fig1} displays three possible 
radial Scarf II potential with origin corresponding to various 
values of $x=r_0$. 

The general solution 
\begin{equation}
\psi(r)=F_1(x)+C F_2(x)
\label{gensol}
\end{equation}
should vanish at $x=r_0$, which defines the constant $C$ as 
\begin{equation}
C=-F_1(r_0)/F_2(r_0)\ .
\label{orig}
\end{equation}
The asymptotic behavior of the solution has to be inspected only for 
$r\rightarrow \infty$, and the $S$-matrix for $l=0$ can be obtained from 
\begin{equation}
\lim_{r\rightarrow\infty} \psi(r) =  
\exp(-{\rm i}kr) -S_0(k)\exp({\rm i}kr)\ . 
\label{smatrix}
\end{equation}
Making use of Eqs. (\ref{ab1+}) and (\ref{ab2+}) of the one-dimensional 
problem the $S$-matrix can be constructed as 
\begin{equation}
S_0(k)=-\frac{a_{1+}+C a_{2+}}{b_{1+}+C b_{2+}}
\label{scarfiism1}
\end{equation}
After some straightforward algebra one obtains 
\begin{eqnarray}
S_0(k)=&&-2^{-4{\rm i}k/c} \exp(\pi k/c) 
\frac{\Gamma(2{\rm i}k/c)}{\Gamma(-2{\rm i}k/c)}
\nonumber \\
&&\times 
\left[
\frac{\Gamma({\rm i}\lambda+1/2-s)}{
 \Gamma({\rm i}\lambda+1/2+{\rm i}k/c)\Gamma(-s+{\rm i}k/c)}
+{\rm i}C 
\frac{\Gamma(-{\rm i}\lambda+3/2+s)\exp(\pi(\lambda+{\rm i}s))}{
 \Gamma(s+1+{\rm i}k/c)\Gamma(-{\rm i}\lambda+1/2+{\rm i}k/c)}
\right]
\nonumber \\
&&\times 
\left[
\frac{\Gamma({\rm i}\lambda+1/2-s)}{
 \Gamma({\rm i}\lambda+1/2-{\rm i}k/c)\Gamma(-s-{\rm i}k/c)}
+{\rm i}C 
\frac{\Gamma(-{\rm i}\lambda+3/2+s)\exp(\pi(\lambda+{\rm i}s))}{
 \Gamma(s+1-{\rm i}k/c)\Gamma(-{\rm i}\lambda+1/2-{\rm i}k/c)}
 \right]^{-1}
\nonumber\\
\label{scarfiism}
\end{eqnarray}
where 
\begin{equation}
C=-\frac{(1+{\rm i}\sinh(cr_0))^{-s+{\rm i}\lambda-1/2}\  
 _2F_1(-s-{\rm i}k/c, -s+{\rm i}k; {\rm i}\lambda-s+1/2; (1+{\rm i}\sinh(cr_0))/2)
}{A 
 _2F_1(1/2-{\rm i}\lambda-{\rm i}k/c, 1/2-{\rm i}\lambda+{\rm i}k/c; 
  s+3/2-{\rm i}\lambda; (1+{\rm i}\sinh(cr_0))/2)
} 
\\
\label{cconst}
\end{equation}

The $S$-matrix of Ref. \cite{npa96} is recovered in the special case of 
$c=1$, 
$\lambda=0$ and $r_0=0$. In that case the radial wave functions are obtained 
from the odd-$n$ solutions of the one-dimensional problem that vanish 
at the origin. 

\begin{figure}[h]
\begin{center}
\resizebox*{10cm}{!}{\includegraphics{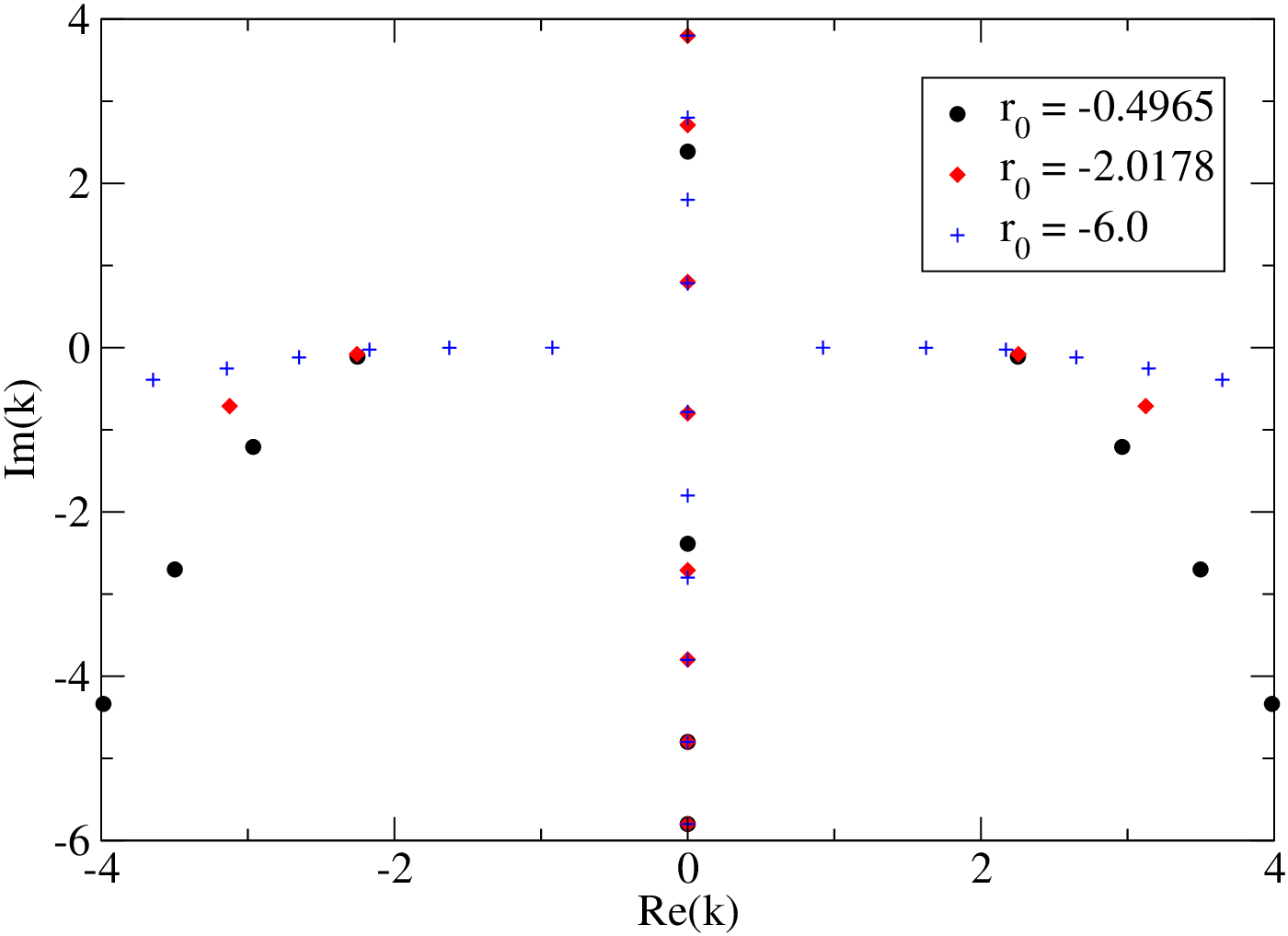}}
\end{center}
\caption{The poles of $S_0(k)$ for the parameters used in Fig. \ref{fig1}.  
Poles located on the positive and negative imaginary axis correspond 
to bound and anti-bound states, respectively, while those appearing 
symmetrically in the third and fourth quadrant represent resonance states. 
}
\label{2-9}
\end{figure}

The poles of the $S$-matrix displayed for the various $r_0$ used in 
Fig. \ref{fig1} are shown in Fig. \ref{2-9}. 

\begin{figure}[h]
\begin{center}
\resizebox*{10cm}{!}{\includegraphics{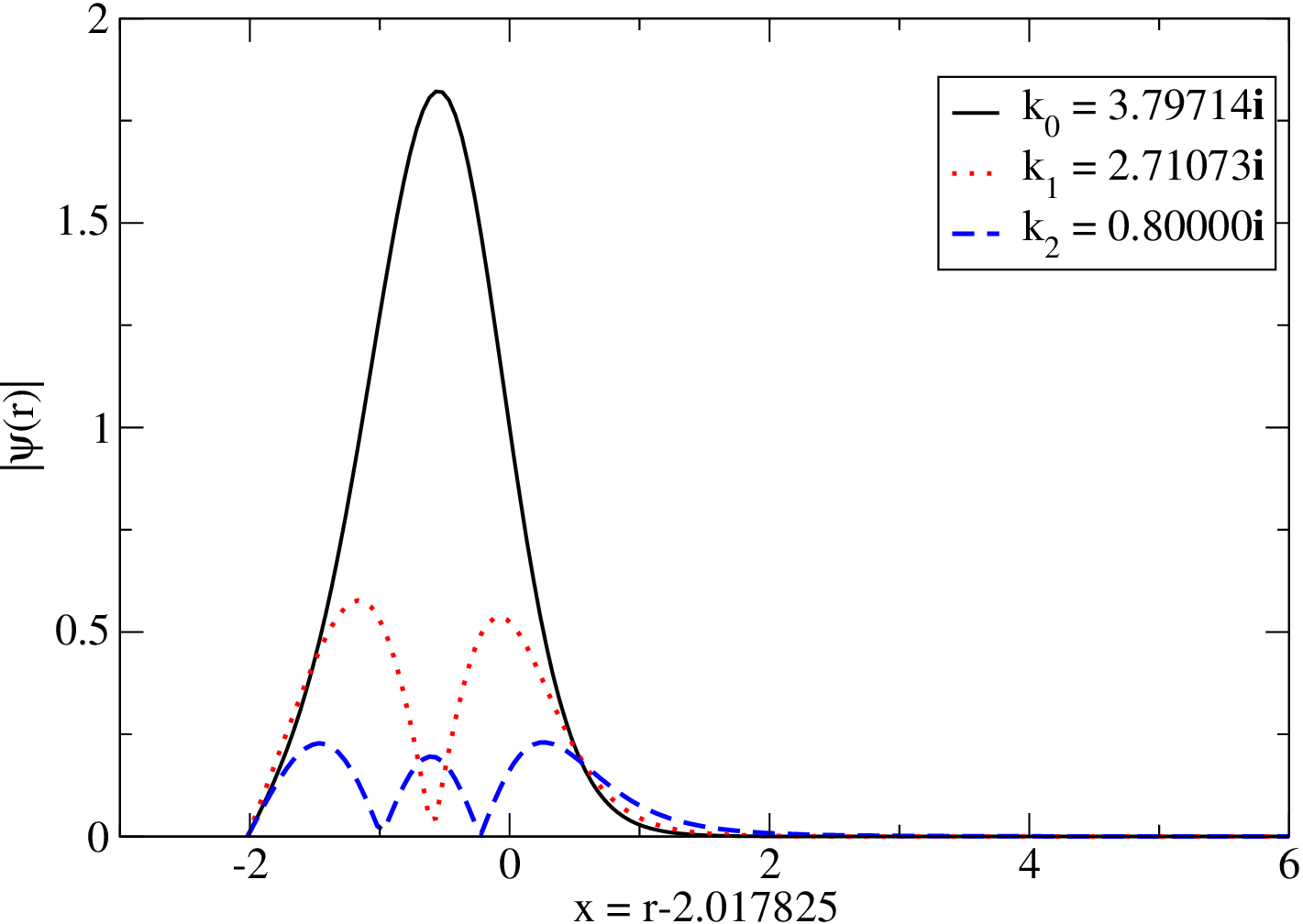}}
\end{center}
\caption{Modulus of the unnormalized $l=0$ bound-state wave 
functions of the radial Scarf II potential for $r_0=-2.017825$.  
The remaining parameters are the same as those used in 
Fig. \ref{fig1}. The second ($n=2$) excited wave function with 
two nodes coincides with the wave function of the one-dimensional 
problem belonging to the same energy eigenvalue, $-0.64$ (see 
Table 1). For the 
normalization of the wave functions, integrals containing the 
product of hypergeometric functions defined on 
$x\in[r_0,\infty)$ would have to be calculated using numerical 
methods.}
\label{fig3}
\end{figure}

The solutions of the one-dimensional and the radial problems can be 
related to each other by various ways. First, if $r_0$ is defined 
to be at a node of a particular wave function $\psi_n(x)$ of the 
one-dimensional problem, then Eq. (\ref{gensol}) implies that 
$\psi_n(r_0)=0$ can occur only for $C=0$, i.e. the solution of 
the radial problem will be the corresponding solution of the 
one-dimensional problem, defined for $x\ge r_0$. Furthermore, the 
energy eigenvalues (and $k$)  will also be the same. This scenario 
is illustrated by the $E_3$ excited state of the one-dimensional 
problem in Table 
1: $r_0=-2.017825$ coincides with the 
first of the three nodes of $\psi_3(x)$ in (\ref{sciiwf}), so 
this function will also act as the second excited state ($n=2$) 
of the radial problem with the same energy eigenvalue 
($E_2=-0.64$), since it has two more nodes. 
The unnormalized bound-state wave functions of this potential 
are displayed in Fig. \ref{fig3}. Obviously, the remaining 
solutions of the radial problem cannot be calculated in the same 
way. The situation is analogous to the case of the harmonic 
oscillator, where some solutions of the radial problem can be 
generated from those of the one-dimensional problem. 

Another relation follows in situations when $r_0$ is defined at 
a large enough negative value, where the bound-state wave functions  
of the one-dimensional problem are close to zero. In this case 
the boundary condition at $r_0$ implies that the second term 
of (\ref{gensol}) should also be small in magnitude ($C$ will be 
small), so the solution will be dominated by $F_1(x)$, i.e. the 
bound-state 
solution of the one-dimensional problem. This also means that 
the energy eigenvalues of the radial problem will also be close 
to those of the one-dimensional problem. A simple test for 
some parameters is displayed in Table \ref{calcener}. 
It is seen that the energy eigenvalues match reasonably well, 
and the agreement gets better with $r_0\rightarrow-\infty$ and 
for lower values of the $n$ quantum number, i.e. in situations 
when the magnitude of $\psi_n(r_0)$ is smaller. 

Considering 
that the resonance solutions do not vanish asymptotically, the 
same argumentation cannot be applied to them. Consequently, the 
resonance energies of the one-dimensional and the radial Scarf II 
potential differ from each other significantly.

\section{Relation to the generalized Woods--Saxon potential and possible 
applications}
\label{gwsanalogy}

It can be noted that the radial Scarf II potential can be brought to a 
form that is close to the notation of the generalized Woods--Saxon potential. 
Applying the variable transformation 
\begin{equation}
cx=c(r-R)\equiv \frac{r-R}{2a}
\label{vartr}
\end{equation}
in (\ref{sciipot}), 
the equivalent form 
\begin{eqnarray}
V(x)&=&-4V_1\frac{\exp((r-R)/a)}{[1+\exp((r-R)/a)]^2}
\nonumber\\
&&+2V_2\left[\frac{\exp((r-R)/(2a))}{1+\exp((r-R)/a)} 
 -2\frac{\exp((r-R)/(2a))}{[1+\exp((r-R)/a)]^2} \right]   
\label{sciipottr}
\end{eqnarray}
is obtained. 
$r_0$ corresponds to $-R$, so the 
radial version of the potential can be obtained by defining the 
origin at the negative value of $r_0=-R$. 

In fact, the same variable transformation relates the generalized 
Woods--Saxon potential with the Rosen--Morse II potential \cite{jpa09b} 
\begin{equation}
V(x)=-\frac{U_1}{\cosh^2(cx)}+U_2\tanh(cx)\ .    
\label{rmiipot}
\end{equation}
After applying the (\ref{vartr}) transformation one obtains  
\begin{equation}
V(r)=
-4U_1 \frac{\exp((r-R)/(a))}{[1+\exp((r-R)/a)]^2}   
-U_2\frac{2}{1+\exp((r-R)/a)} +U_2\ ,
\label{rmiipotws}
\end{equation}
which is the Woods--Saxon potential with a shifted energy scale.  
Note that the first terms of (\ref{sciipottr}) and (\ref{rmiipotws}) 
are the same, and also that the diffusity parameter $a$ is related 
to $c$ via (\ref{vartr}). 
The first ``surface'' term of (\ref{rmiipotws}) (and thus of 
(\ref{rmiipot})) can be obtained as the first derivative of second 
``volume'' term. There is no similar relation for the 
Scarf II potential: however, the second term of (\ref{sciipot}) 
can be obtained from the first-order derivative of 
$(\cosh(cx))^{-1}$, i.e. the square root of the first term. 
Obviously, similar relation holds for (\ref{sciipottr}) too.  

It should be noted that the Rosen--Morse II potential is defined 
on the full real $x$ axis, so in order to obtain the Woods--Saxon 
potential from it, one should consider $r\in [0,\infty)$ in 
(\ref{vartr}). This means that the two solutions have to be 
matched at $r=0$ in a way similar to that considered previously for 
the Scarf II potential. In fact, the procedure applied there is the 
same as that described in the notable work of Bencze \cite{bencze}, 
where the $S$ matrix of the Woods--Saxon potential was constructed 
in an analytical form. See also Ref. \cite{npa16} as a more 
accessible source of the formulas. 

Note that in the one-dimensional Rosen--Morse II potential the 
normalizable states are expressed in terms of only one of the 
two independent solutions (similarly to the case of the one-dimensional 
Scarf II potential), and they exist only for $U_1>0$ \cite{jpa09b}. 
In contrast with this, in the radial problem, this ``surface'' term usually 
plays the role of a barrier, i.e. $U_1<0$, and the attractive component 
of the potential is represented by the ``volume'' term with $U_2>0$.  
The relation between the Rosen--Morse II and the generalized 
Woods--Saxon potential has been pointed out in Ref. \cite{berkd}, 
where analytical expressions were given for the $l=0$ bound-state 
wave functions  and the corresponding energy eigenvalues. However, 
those wave functions do not vanish at the origin (their structure 
is similar to the wave functions of the one-dimensional Rosen--Morse II 
potential), so that approach can be considered as an approximation 
only. 

Given the analogy with the generalized Woods--Saxon potential, 
possible applications of the radial Scarf II potential can be envisaged 
in nuclear physics, for example. The barrier in Fig. \ref{fig1} can 
simulate the effects of the Coulomb barrier that occurs when charged 
particles (e.g. protons or $\alpha$-particles) interact with a nucleus. 
The situation is qualitatively similar to the case of the generalized 
Woods--Saxon potential, which has a similar barrier, and the difference 
occurs within the nucleus, where the latter potential is constant, 
while the radial Scarf II potential has a clear minimum, and 
depending on $r_0$ it can increase close to the origin. 
Potentials of this kind might be relevant to nuclei exhibiting 
depleted central nucleon density. Such ``bubble'' nuclei have been 
observed in recent experiments \cite{bubble}. 

It is also possible to shift the barrier {\it inside} the nucleus 
by formally reflecting the potential curve in Fig. \ref{fig1} 
about $x=0$ and defining the origin near the potential maximum. 
Staying with the original formalism, this corresponds to considering 
$V_2<0$, i.e. taking $\alpha\leftrightarrow\beta$ in (\ref{v1v2}) or 
$\lambda\rightarrow -\lambda$ in (\ref{v1v2sl}). 
Potentials with such shape occur in 
hypernuclei, where the interaction of $\Lambda$ particles with $\alpha$ 
particles or nucleons requires the presence of a soft repulsive core 
with variable heigth \cite{daska}. 

Finally, the formalism can be extended to the case of optical potentials 
with complex values of $V_1$ and $V_2$: the calculation of the 
wave functions, $S$-matrix, energy eigenvalues, including that of 
the resonances will be the same.

\section{Summary}
\label{summary}

Based on the results of the one-dimensional Scarf II potential, the 
radial version of this potential was studied. For this, the origin 
was defined at an arbitrary value on the real $x$ axis, and the 
$s$-wave solutions were constructed from the two independent 
solutions of the one-dimensional Schr\"odinger equation, after 
prescribing the appropriate boundary conditions. The asymptotic 
form of these solutions was used to construct the $S_0(k)$ 
$S$-matrix. The poles of $S_0(k)$ were located, and were 
identified with the bound, anti-bound and resonance solutions. 

It was shown that by selecting the origin far enough from 
the potential minimum, the bound-state energy eigenvalues and 
wave functions of the radial potential tended to those of the 
one-dimensional potential. Furthermore, selecting the origin 
at the node of some bound-state wave function of the one-dimensional 
potential, this wave function appeared as a bound-state wave 
function of the radial potential with the same energy eigenvalue. 

With a slightly modified parametrization, the radial Scarf II 
potential could be compared with the generalized Woods--Saxon  
potential, and it was shown that they share a term (the ``surface'' 
term of the latter potential). In fact, it was demonstrated that 
the radial Scarf II potential can be generated from the 
one-dimensional Scarf II potential in the same way as the 
generalized Woods--Saxon 
potential is generated from the one-dimensional Rosen--Morse II  
potential. The connection between the latter two potentials has 
been known before \cite{berkd}, however, the bound-state wave 
functions generated from this connection did not satisfy the 
appropriate boundary conditions. 

Based on its similarity with the generalized Woods--Saxon potential, 
the radial Scarf II potential could be applied in nuclear physics, 
for example. 
One possibility is considering problems, which are characterized 
by a barrier at the surface of the nucleus, but in which the flat 
potential inside the nucleus is replaced with a potential well 
with a clear minimum. Another option is placing the barrier inside 
the nucleus near the origin, simulating a repulsive interaction 
there. The formalism can be extended to the case of complex values 
of $V_1$ and $V_2$, i.e. to optical potentials.

\section*{Acknowledgments}
This work was supported by the Hungarian Scientific Research 
Fund -- OTKA, grant No. K112962.









\begin{table}[h!]
\begin{center}
\caption{Bound-state energies of the one-dimensional Scarf II 
potential and those of the radial one defined with various $r_0$ 
in Fig. \ref{fig1}.}
\label{calcener}
\begin{tabular}{lcccc}
\hline
 & 1D case & $r_0=-0.4965$ & $r_0=-2.017825$ & $r_0=-6.0$ \\
\hline\\
 $E_0$ & $-14.44$ & $-5.69802$ & $-14.41825$ & $-14.44000$ \\
$E_1$ & $-7.84$ & -- & $-7.34803$ & $-7.84000$ \\
$E_2$ & $-3.24$ & -- & $-0.64000$ & $-3.23998$ \\
$E_3$ & $-0.64$ & -- & -- & $-0.61366$ \\
\hline
\end{tabular}
\end{center}
\label{tab1}
\end{table}

\end{document}